\documentclass[journal,12pt,onecolumn]{IEEEtran}
\usepackage{cite,url,enumerate}
\usepackage[cmex10]{amsmath}
\usepackage{amssymb,amsthm}

 \newcommand\bP{\mathbb P} 
\newcommand{\bbsm}{( \begin{smallmatrix}}      \newcommand{\besm}{\end{smallmatrix} )}
\newcommand{\bbm}{\begin{pmatrix}}      \newcommand{\bem}{\end{pmatrix}}
\newcommand\beq{\begin{equation}}  \newcommand\eeq{\end{equation}}
\newcommand\beqs{\begin{equation*}}  \newcommand\eeqs{\end{equation*}}
\newcommand\bep{\begin{IEEEproof}}  \newcommand\eep{\end{IEEEproof}}

  \newcommand\mC{\mathcal C}
 \newcommand\mD{\mathcal D} \newcommand\mO{\mathcal O}
 \newcommand\mG{\mathcal G} 
  \newcommand\GFm{GF(q)^{\times}}

\newtheorem{theorem}{Theorem}
\newtheorem{lemma}{Lemma}
\newtheorem{proposition}{Proposition}
\newtheorem{corollary}{Corollary}
\newtheorem{conjecture}{Conjecture}
\newtheorem{definition}{Definition}

\newtheorem*{conjecture*}{Conjecture}
\newtheorem*{theorem*}{Theorem}

\newenvironment{remark}[1][Remark]{\begin{trivlist}
\item[\hskip \labelsep {\bfseries #1}]}{\end{trivlist}}

\begin{document}
\title{Deep holes  and MDS extensions of Reed-Solomon codes.}
\author{Krishna~Kaipa
\thanks{K. Kaipa is with the Department of Mathematics at the  Indian Institute of Science and Education Research, Pune  411008, India (email: kaipa@iiserpune.ac.in).}}
\maketitle
\begin{abstract}
We study the problem of classifying  deep holes of Reed-Solomon codes.   We show that this problem is equivalent to the problem of classifying MDS extensions 
of Reed-Solomon codes by one digit.  This equivalence allows us to improve  recent results on the former problem.
In particular, we classify deep holes of Reed-Solomon codes  of dimension  greater than half the alphabet size. 
 We also give a complete classification of deep holes of Reed Solomon codes with redundancy three in all dimensions.  
\end{abstract}

\IEEEpeerreviewmaketitle

\section{Introduction} \label{i}
A deep-hole for a code $\mC$ is a received vector whose distance from $\mC$  attains the  maximum possible value viz. the covering radius of $\mC$. 
A  $[n,k, \mD]_q$  RS (Reed-Solomon) code $\mC$ consists of codewords  $(f(x_1),  f(x_2), \dots, f(x_n))$  as $f(X)$ runs over  the set of  univariate polynomials of degree at most $k-1$ with $GF(q)$-coefficients. The evaluation set   $\mD = \{x_1, \dots, x_n\}$ consists of $n$ distinct and ordered points of $GF(q)$.
  The covering radius of $\mC$ can be shown  to be $n-k$. 
By a generating polynomial $u(x)$ for a received word  $u \in GF(q)^n$ of $\mC$, we mean the Lagrange interpolation polynomial of degree at most $n-1$ for the data points  $\{(x_1, u_1), \dots, (x_n, u_n)\}$.  It was shown in  \cite{Gu_Va} that the problem of  determining  whether a received word is a deep hole of a given Reed Solomon code is NP-hard.
Several authors have studied the problem of classifying deep holes of RS codes. Cheng and Murray conjectured that:
\begin{conjecture} [Cheng and Murray \cite{Ch_Mu}]  \label{conj1}   All deep-holes  of a $[k,q, \mD=GF(q)]_q$ RS code  with $2 \leq k \leq q-2
$ have generating polynomials of degree $k$, except when $q$ is even and $k=q-3$.
\end{conjecture}

The exception was not part of \cite{Ch_Mu}, but is well known. In fact this conjecture is equivalent to a central conjecture in finite geometry (See conjecture 1). 
It may be somewhat surprising to note that the conjecture for $k=q-3$ and $q$ odd, is equivalent to Beniamino Segre's  foundational theorem of finite geometry
 that states -- in coding theory terminology-- that any $3$-dimensional MDS code of length $q+1$ is Reed-Solomon. Recently,  Zhuang, Cheng and Li obtained the following result:
\begin{theorem*}  [Zhuang, Cheng and Li \cite{ZCL}]  \label{ZCL_thm}  Let $\mC$ be a $[n,k, \mD]_q$ RS code  with $k \geq   \lfloor (q-1)/2 \rfloor$. If $q > 2$ is a prime number and  then the generating polynomials of  deep holes of  $\mC$  are 
\[  u(x)=a u_1(X)  + u_2(X),  \; a \in \GFm, \,  \text{deg}(u_2) \leq k-1,\]  and $u_1(X)$ equals either  $X^k$ or the 
generating polynomial for the data points $\{(x_i, \tfrac{1}{x_i - \delta}) :1 \leq i  \leq n\}$  for some $\delta \in GF(q)\!\setminus\!\mD$. \\
\end{theorem*}

The techniques used in \cite{ZCL} necessitate the condition that the alphabet size is an odd prime, and they leave the problem open for $GF(q)$. Our method (which uses
the work of  Roth and Seroussi \cite{Ro_Se}) allows us to work with any finite field  alphabet. The contributions of this work are as follows :
\begin{theorem}    \label{main_thm} \hfill \begin{enumerate} \item  The above theorem  of Zhang et al. holds not only for $GF(p)$ but for any  $GF(q)$ with $q$ odd. 
\item For $q$ even,  the result of the above theorem  still holds except when $n=k+3$. 
In this  case
$u_1(X)$ may also equal $X^{k+1} + (\sum_{i=1}^{n} x_i) X^{k}$ in addition to choices for $u_1(X)$ mentioned therein.
\end{enumerate}
\end{theorem}
 In Theorem \ref{geom_thm},  we give a geometric interpretation of  Theorem \ref{main_thm}.
Our method uses the observation that for a $[n,k,\mD]$ RS code with ($n<q$) and a parity check matrix $H$ of $\mC$,  a received vector $u \in GF(q)^n$ is a deep hole of $\mC$ if and only if the syndrome $Hu$ has the property that the matrix $[H \,|\, Hu]$ is an MDS extension of the RS code $\mC^{\perp}$ by one digit. In the   case $k \geq   \lfloor (q-1)/2 \rfloor$, 
the results of Roth and Seroussi \cite{Ro_Se}, allow us to obtain Theorem \ref{main_thm}.\\

For $[n,k]_q$ RS codes $\mC$ of length $q+1$ (here the evaluation set is $\mD = PG(1,q)$), it is not always true that the covering radius of $\mC$ is $n-k$. Sometimes the covering radius equals $n-k-1$. In the latter case the equivalence between deep holes of $\mC$ and MDS extensions of $\mC^{\perp}$ breaks down. 
In Section \ref{gdrs}, we present some  results for RS codes of length $q+1$. \\

In Section \ref{red3}  we  classify  deep-holes of a $[n,k]_q$ RS codes with redundancy $n-k=3$ (Theorem \ref{red3_thm}). We also obtain canonical forms for non GRS $3$-dimensional MDS codes extending a GRS code by one digit. (Theorem \ref{thm_canon}).

\section{Notation} \label{ii}
 Throughout this article, the dimension $k$ of a RS code is in the range $2 \leq k \leq q-1$.
As for the length $n$, we note that the cases where the length $n$ equals $k$ or $k+1$ are uninteresting:  In the former case there are no deep holes, and in the later case deep-holes are just those received words which are not codewords. Therefore, we impose the condition  $k+2 \leq n \leq q+1$ on the length of $\mC$.\\

 Generalized Reed Solomon codes (GRS codes) are obtained from RS
codes by applying a diagonal Hamming isometry $u  \mapsto  \text{diag}(\nu_1, \dots, \nu_n) u$ of $GF(q)^n$. Clearly the deep holes of the resulting GRS code are obtained 
from those of the original RS code by applying the same transformation.  Thus, for the purposes of studying deep holes, it is sufficient to consider only RS codes.
The dual of a RS code, is   a GRS code.  The definition of a $[n,k, \mD]_q$ RS code is easily extended to allow the evaluation set $\mD$ to be a subset of the projective line $PG(1,q) = GF(q) \cup \infty$. Here it is understood that the value of a message polynomial $f(X)$ at $\infty$ is the coefficient of $X^{k-1}$.   The evaluation set $\mD$ of a GRS code is not unique: if $\varphi(x)= (c+dx)/(a+bx)$ is a fractional linear transformation of  $PG(1,q)$ then $\varphi(\mD)$ is also an evaluation set (for example see Proposition 2.5 of \cite{BGHK}). Using this freedom we can (and will) always assume $\mD$ to be free of $\infty$ provided $n \neq q+1$.\\

Given an evaluation set $\mD = \{x_1, \dots,x_n\} \subset GF(q)$, we associate to $\mD$ some numbers $s_0, \dots, s_n$ and $\nu_1, \dots, \nu_n$  defined by:
\beq  \label{eq:nu}   \prod_{i=1}^n (X - x_i) = \sum_{j=0}^n s_j X^{n-j}, \quad \nu_j =\prod_{\{i: i \neq j\}} (x_j - x_i)^{-1}.\eeq
 In order to give geometric interpretation to our results, it will be convenient to introduce some terminology. We recall that an $[n,k]_q$ MDS code is a $k$-dimensional code of length $n$ whose minimum distance is $n-k+1$. Equivalently any generator matrix of a $[n,k]_q$ MDS code has the property that all its $k \times k$ minors are non-zero.  An $n$-arc in $PG(k-1,q)$ is a unordered collection of $n$ points  of $PG(k-1,q)$ such that any $k \times n$ matrix whose columns are lifts of the $n$ points of the arc to  $GF(q)^n$,  generates a $[n,k]_q$ MDS code.   Thus there is a  bijective correspondence between monomial equivalence classes of $[n,k]_q$ MDS codes and  projective equivalence classes of $n$-arcs in $PG(k-1,q)$. For each element $x \in GF(q) \cup \infty$  we define vectors $c_m(x) \in GF(q)^m$ by:
  \beq \label{eq:rnc} c_{m}(x) = \begin{cases} (1,x,x^2, \dots, x^{m-1})^T  &\text{ if } x \in GF(q) \\
 (0,0,\cdots, 0, 1)^T &\text{ if } x = \infty. \end{cases} \eeq 
We note that a $[n,k,\mD]_q$ RS code with $\mD = \{x_1, \dots, x_n\} \subset GF(q)$ has a generator and parity check matrix pair  given by:
\begin{IEEEeqnarray}{rCl} \label{eq:gen_GRS} 
 G_k(\mD) &=&    [c_k(x_1)  \,| \,  c_k(x_2)  \,| \, \dots  \,| \, c_k(x_n)] \\
\nonumber G_k^{\perp}(\mD) &=& [\nu_1 c_{n-k}(x_1)  \,| \, \nu_2 c_{n-k}(x_2)  \,| \, \dots  \,| \, \nu_n  c_{n-k}k(x_n)]
\end{IEEEeqnarray}
where $\nu_i$ are as in \eqref{eq:nu}.
In the case $\mD = \{x_1, \dots,x_q\} \cup \infty$,  the $[q+1,k,\mD]$ RS code $\mC$ has a generator and parity check matrix pair  given by:
\begin{IEEEeqnarray}{rCl}   \label{eq:gen_GDRS} 
G_k(PG(1,q)) &=& [c_k(x_1)  \,| \,  c_k(x_2)  \,| \, \dots  \,| \, c_k(x_q) \,| \, c_k(\infty)] \\
\nonumber G_k^{\perp}(PG(1,q)) &=& [c_{q+1-k}(x_1)  \,| \,  c_{q+1-k}(x_2)  \,| \, \dots  \,| \,  c_{q+1-k}(x_q) \,|\, c_{q+1-k}(\infty)].
\end{IEEEeqnarray}

The standard RNC (rational normal curve) in $PG(m-1,q)$ consists of $q+1$ points of $PG(m-1,q)$ given  by $\{[c_m(x)] : x \in GF(q) \cup \infty\}$.
(For a nonzero  $v \in GF(q)^m$, we use the notation $[v] \in PG(m-1,q)$ to denote the one-dimensional subspace of $GF(q)^m$  represented by $v$).
By a RNC in $PG(m-1,q)$, we mean the image of the standard RNC under a projective linear transformation of $PG(m-1,q)$. Thus,  in the correspondence between arcs and MDS codes, the  $n$-arcs of $PG(k-1,q)$ which correspond to $[n,k]_q$ RS codes, are those arcs which are contained in  a RNC. A $n$-arc in $PG(k-1,q)$ is said to be complete if it is not contained in a $n+1$-arc. Equivalently, the corresponding $[n,k]_q$ MDS code cannot be extended to a $[n+1,k]_q$ MDS code. Let  $\mathcal N_m \in GF(q)^m$ denote the vector $(0,\dots,0,1,0)^T$.   For $q$ even, the point $\mathcal N_3 \in PG(2,q)$ is known as the nucleus of the standard RNC in $PG(2,q)$.

\begin{definition} \label{equiv} For a $[n,k]$ code $\mC$, we say two received words $u_1, u_2 \in GF(q)^n$ are coset-equivalent if $u_2 - u_1 \in \mC$. We say $u_1, u_2$ are equivalent if $u_2 - a u_1 \in \mC$ for some  $a \in \GFm$. 
\end{definition}
(In this artcle we denote $GF(q) \setminus \{0\}$ by $\GFm$).
In the case $\mD \subset GF(q)$, if $u(X)$ is the generating polynomial of a deep hole $u$, then the generating polynomials of words equivalent to $u$ are $\{ a u(X) +f(X) : a \neq 0, \, \text{deg}(f) <k\}$. Thus there is a unique $v$ equivalent to $u$  whose generating polynomial is of the form $X^k P_{u}(X)$ with $P_u(X)$ monic of degree at most $n-k-1$. 

We use the notation $\rho(\mC)$ for the covering radius of a code $\mC$. 

\section{Deep holes and MDS extensions} \label{mdsext}
Let $\mC$ be a $[n,k,\mD]$ RS code.  If $n \neq q+1$,  we recall the proof that the covering radius $\rho(\mC)=n-k$. Since  $\rho(\mC) \leq n-k$  for any linear $[n,k]_q$ code, we just need to show there is a received word at a distance of $n-k$ from $\mC$. The word $(x_1^k, \dots, x_n^k)$ is at a distance  $n-k$ from $\mC$: 
the vector  $(p(x_1)-x_1^k, \dots, p(x_n)-x_n^k)$  for deg$(p(X)) <k$ has at most $k$ zeros, and there is a $p(X)$ for which this vector has $k$ zeros.\\

In the case $n=q+1$, let $\mD = \{x_1, \dots, x_q, \infty\}$ with $\{x_1, \dots, x_q\}= GF(q)$. Consider a word of the form $u=(x_1^k, \dots,x_q^k,a)$ for any $a \in GF(q)$.
We will show that the distance of $u$ from $\mC$ is $n-k-1$. Let $k \, {}^\wedge GF(q)$ denote the subset of $GF(q)$ consists of those elements which can be written as a sum of $k$ distinct elements of $GF(q)$. It is easy to see that  $ c \cdot (k \, {}^\wedge GF(q)) = k \, {}^\wedge GF(q)$ for all $c \in \GFm$ and hence $k \, {}^\wedge GF(q) = GF(q)$. Thus, there exist distinct elements $x_1, \dots, x_k \in GF(q)$   such that $a = x_1+ \dots+x_k$. Consider the polynomial $p(X) = X^k - \prod_{i=1}^k (X - x_i)$. This is a codeword (message word) of $\mC$ and the vector 
$(p(x_1)-x_1^k, \dots, p(x_q)-x_q^k, p(\infty) - a)$ has exactly $k+1$ zeros. This is because $p(\infty) = \sum_{i=1}^k x_i =  a$. Thus, we conclude that $\rho(\mC)$ is either $n-k-1$ or $n-k$. We will see in Section \ref{gdrs} that both situations occur. \\

Let $G_k(\mD)$ and $G_k^{\perp}(\mD)$ be a generator and parity check matrix for $\mC$ as given in \eqref{eq:gen_GRS} if $n \neq q+1$ and 
\eqref{eq:gen_GDRS}  if $n=q+1$. For a received word $u$ let $S_{\mD}(u) = G_k^{\perp}(\mD) u$ be the syndrome of $u$. When $u$ is a non-codeword, 
we  use the term projective syndrome for $[S_{\mD}(u)] \in PG(n-k-1,q)$.
\begin{proposition} \label{main_prop}
  Let $\mC$   be  a $[n,k,\mD]_q$ RS code. In case $n=q+1$,  suppose $\rho(\mC)=n-k$.\\ For a received word $u \in GF(q)^n$, the augmented matrix:
\beq \label{eq:gen_ext}  G_k^{\perp}(\mD;u):= [ G_k^{\perp}(\mD) \, | \, S_{\mD}(u) ] \eeq
generates a $[n+1,n-k]_q$ MDS code  if and only if  $u$  is deep-hole of $\mC$. 
Thus $S_{\mD}$  sets up  a bijective  correspondence between the set  of coset-equivalence classes   of  deep-holes  and the set of MDS extensions  of the dual RS code $\mC^{\perp}$  by one digit. 
\end{proposition}
\bep
We recall that the distance of $u$ from $\mC$ is the least number $m$  such that $S_{\mD}(u)$ can be written as a linear combination of some $m$ columns of $G_k^{\perp}(\mD)$. 
Suppose $\rho(\mC)  = n-k$ (this is automatic if $n \neq q+1$). It follows that $u$ is a deep hole of $\mC$ if and only if $[G_k^{\perp}(\mD)\,|\, S_{\mD}(u) ]$ generates an $[n+1,k]$ MDS code extending the GRS code $\mC^{\perp}$.  The second assertion follows from the fact that two  received words  are coset equivalent if and only if their syndromes coincide. 
\eep

We need the following result:
\begin{theorem*} [Roth and Seroussi 1986  \cite{Ro_Se}] \label{ro_se_thm} Let $\mC$ be a $[n, \ell,\mD]$ RS code. Suppose $2 \leq \ell \leq n - \lfloor (q-1)/2 \rfloor$. Let $g \in GF(q)^{\ell}$ be a vector. The augmented matrix $[ G_{\ell}(\mD) \,|\, g]$ generates an $[n+1,\ell]_q$ MDS code if and only if:
\begin{enumerate}
\item (for $q$ odd) $[g]=[c_{\ell}(\delta)]$ for some $\delta \in PG(1,q) \!\setminus\! \mD$
\item (for $q$ even) $g$ is either as above, or additionally in case $\ell=3$, $[g] = \mathcal N_3$. 
\end{enumerate}
\end{theorem*}

Under the hypothesis of Proposition \ref{main_prop} and  $k \geq \lfloor (q-1)/2 \rfloor$, the matrix $G_k^{\perp}(\mD;u)$ in  \eqref{eq:gen_ext} generates a MDS code,  if and only if $g=S_{\mD}(u)$  has the form given in the above theorem with $\ell = n-k$.
Thus we have proved:

\begin{theorem}  [Geometric form of Theorem \ref{main_thm}] \label{geom_thm}
Under the hypothesis of Proposition \ref{main_prop} and and  $k \geq \lfloor (q-1)/2 \rfloor$, a received word $u$ is a deep hole of $\mC$ if and only if: \begin{enumerate}[a)]
\item  either $[S_{\mD}(u)] = [c_{n-k}(\delta)]$ lies on the standard RNC in $PG(n-k-1,q)$ for some $\delta \in PG(1,q) \setminus \mD$.
\item  or  $q$ is even, $n=k+3$, and  $[S_{\mD}(u)]$ is as in part a)  or equals the nucleus $\mathcal N_3$  of the standard $RNC$ in $PG(2,q)$.
\end{enumerate}
\end{theorem}
Before, we prove Theorem \ref{main_thm}, we need some lemmas.
We now assume $\mC$ is a $[n,k,\mD]$ RS-code with $\mD \subset GF(q)$. Given a vector $u \in GF(q)^n$, let 
 $u(X)$ be   the generating polynomial of $u$. Clearly $u \mapsto u(X)$ is a linear isomorphism:
\begin{lemma} A formula for $u(X)$ in terms of $u$, $s_0, \dots ,s_n$  and $\nu_1, \dots, \nu_n$ is:
\beq \label{eq:u*}  u(X) = [X^{n-1}, X^{n-2}, \dots, X, 1] L_n \,  [\nu_1 c_n(x_1)  \,| \,  \nu_2 c_n(x_2)  \,| \, \dots  \,| \, \nu_n c_n(x_n)]\, u \eeq
where $L_n$ is  the $n \times n$ lower triangular matrix  given by $L_{ij} = s_{i-j}$.\\
Moreover, writing $u(X) = u_1(X) + u_2(X)$ where deg$(u_2) <k$ and $u_1$ only contains monomials $X^i$ for $i \geq k$, we see that
\beq \label{eq:u_1}  u_1(X) = [X^{n-1}, X^{n-2}, \dots, X^k] L_{n-k} S_{\mD}(u), \eeq
where $L_{n-k}$ is the submatrix of $L_n$ on the first $n-k$ rows and columns.
\end{lemma}
\bep The right hand side of \eqref{eq:u*} simplifies to the Lagrange interpolation polynomial 
\[ u(X)= \sum_{j=1}^n u_j \nu_j \prod_{\{i: i \neq j\}} (X - x_i) \]
Clearly, 
\[ u_1(X)=[X^{n-1}, X^{n-2}, \dots, X^k]  L_{n-k} \, G_{k}^{\perp}(\mD) u \]
which is the same as the formula in \eqref{eq:u_1}.
\eep
\begin{lemma} \label{lem_del} Let $u = (u_1, \dots, u_n)$ be the vector with $u_i = 1/(x_i  - \delta)$ where $\delta \in GF(q) \setminus \mD$. The generating polynomial of $u$ is 
\beq \label{eq:delpol} u(X)= a\,  [X^{n-1}, X^{n-2}, \dots, X, 1] L_n \, c_n(\delta) \eeq
where $a = -1/ \prod_{i=1}^n (\delta - x_i)$.
\end{lemma}
\bep By  Lagrange interpolation
\[ u(X) = \prod_{i=1}^n    \nu_i (x_i - \delta)^{-1} \prod_{\{\mu : \mu \neq i\}}  (X - x_{\mu}). \]
From the definition of the quantitities $\nu_i$, it follows that $(X - \delta) u(X)$ is the  Lagrange interpolation polynomial  of degree at most $n$ for the data $ \{(x,1) : x \in \mD\} \cup \{(\delta,0)\}$. In other words
\[ (X - \delta) u(X) = 1 +a \prod_{i=1}^n (X-x_i)= 1+a  \sum_{j=0}^n s_j X^{n-j}. \]
Using this equation to determine the coefficients of $u(X)$, we obtain 
\[u(X) = a  \sum_{j=1}^n  \, X^{n-j}  (\sum_{i=0}^{j-1} \delta^{j-i-1} s_i),  \]
which is the same as  the formula \eqref{eq:delpol}
\eep

 \bep \emph{(of Theorem \ref{main_thm})}  According to Theorem \ref{geom_thm}, for  $k \geq \lfloor (q-1)/2 \rfloor$, and $n \neq k+3$ if $q$ is even,  $[S_{\mD}(u)] = [c_{n-k}(\delta)]$ lies on the standard RNC in $PG(n-k-1,q)$ for some $\delta \in PG(1,q) \setminus \mD$.  When $\delta = \infty$,  $S_{\mD}(u) =a  c_{n-k}(\infty)$ for some $a \in \GFm$, and thus 
$L_{n-k} S_{\mD}(u)$ is $a$ times the last column of $L_{n-k}$ which is again $c_{n-k}(\infty)$. Thus 
the formula \eqref{eq:u_1} implies that $u_1(X) =a  X^k$, as was to be shown. In case $\delta \in GF(q) \setminus \mD$, 
$S_{\mD}(u) =a  c_{n-k}(\delta )$ for some $a \in \GFm$. The generating polynomial $u(X)$ is given by the formula  \eqref{eq:u*}. Comparing with equations \eqref{eq:u_1} and \eqref{eq:delpol} of Lemma \ref{lem_del}, we see that $u(X)$ is of the form   $u(X)=b f(X)  + g(X)$, where    $b \in \GFm$,  deg$(u_2) \leq k-1$,   and $f(X)$ is the
generating polynomial for the data points $\{(x_i, \tfrac{1}{x_i - \delta}) :1 \leq i  \leq n\}$.\\
In the case    $q$ is even, $n=k+3$ and $k \geq \lfloor (q-1)/2 \rfloor$,   and $[S_{\mD}(u)] \neq [c_{n-k}(\delta)]$ for $\delta \in PG(1,q) \setminus \mD$, we must have $[S_{\mD}(u)]= (0:1:0)$  by Theorem \ref{geom_thm}.  It follows from \eqref{eq:u_1} that 
\[ u(X) = a [X^{k+2}, X^{k+1}, X^k] \bbsm 1 &0& 0\\ s_1& 1& 0\\ s_2 &s_1& 1 \besm \, \bbsm 0 \\1 \\0 \besm +u_2(X)= a(X^{k+1} + s_1 X^k) +u_2(X),\]
for some polynomial $u_2$ of degree at most $k-1$.
\eep

We  now give a geometric restatement of  the Conjecture \ref{conj1} of Cheng and Murray that we mentioned in Section \ref{i}.  Since $\mD = GF(q)$, we note that $PG(1,q) \setminus \mD = \{\infty\}$.  To say that the generating polynomial of a deep hole has degree exactly $k$ is equivalent to the assertion that $[S_{\mD}(u)] = [c_{n-k}(\infty)]$ (by \eqref{eq:u_1}). Thus Conjecture \ref{conj1}  is equivalent to the following conjecture (where  $m=q-k$)
\begin{conjecture*}  ${\bf 1'}$ 
For  $2 \leq m \leq q-2$, with the exception of  $m=3$ when $q$ is even,   any  MDS extension (by one digit) of a $m$ dimensional $RS$ code with evaluation set $GF(q)$ must itself be GRS.\\  Equivalently,   any  $(q+1)$-arc in $PG(m-1,q)$ with $q$ points on a RNC must have all its points on the RNC. 
\end{conjecture*}

\begin{corollary} (of Theorem \ref{geom_thm})  The conjecture of Cheng and Murray holds if $k \geq \lfloor (q-1)/2 \rfloor$, except when $q$ is even and $k=q-3$. 
\end{corollary}

\begin{remark}\!: We strongly believe that Conjecture \ref{conj1} must hold for all $q-2 \geq k \geq 2$, not just $k \geq \lfloor (q-1)/2 \rfloor$.  We justify this belief with the next two Propositions. \end{remark}
\begin{proposition} \label{segre_prop}   For $q$ odd, Conjecture  \ref{conj1}  holds for $k=2$  (i.e. Conjecture ${\bf 1'}$ holds for $m = q-2$), as it is equivalent to  B. Segre's fundamental  result \cite{Segre1}  that any $q+1$-arc in $PG(2,q)$ (for odd $q$) is a RNC.  Equivalently, for odd $q$ any $[q+1,3]_q$ MDS code is GRS.
\end{proposition}
\bep Let $u$ be a deep hole of a $[q,2]_q$ RS code $\mC$ with $\mD = GF(q)$.  According to Proposition \ref{main_prop}, 
the matrix $G_2^{\perp}(\mD;u)$ generates a  $[q+1,q-2]_q$ MDS code $\mC_1$ extending the $[q,q-2]_q$ RS code $\mC^{\perp}$. Consider the 
$[q+1,3]_q$ MDS code $\mC_1^{\perp}$. Segre's theorem implies $\mC_1^{\perp}$ is GRS. Since the dual of a GRS code is GRS, it follows that $\mC_1$ is GRS. Thus the columns of the matrix $G_2^{\perp}(\mD;u)$ represent the $q+1$ points of some RNC in $PG(q-3,q)$. Now, a RNC in $PG(m-1,q)$ is uniquely determined by any $m+2$ points on it. (see \cite[Theorem 21.1.1 (v)]{Hirschfeld1985}, or \cite[ Theorem 2.7 ]{BGHK} for a purely coding theoretic proof.) 
Since $m+2 = q$ here, and the first $q$ columns of $G_2^{\perp}(\mD;u)$ lie on the standard RNC in $PG(q-3,q)$, we conclude that the columns of $G_2^{\perp}(\mD;u)$ represent the $q+1$ points of the standard RNC.\\

Conversely, we show Segre's theorem is equivalent to Conjecture \ref{conj1} holding for $k=2$. Given a $q+1$-arc in $PG(2,q)$, or in other words a $[q+1,3]_q$ MDS code $\mC$, the dual code $\mC^{\perp}$ is a $[q+1,q-2]$ MDS code. Puncturing $\mC^{\perp}$ on the last coordinate gives a $[q,q-2]$ MDS code $\mC_1$. The dual to this code is a $[q,2]$ MDS code and $2$-dimensional MDS codes are always GRS. Thus $\mC_1$ is GRS. It follows that $\mC^{\perp}$ is a $[q+1,q-2]$ MDS code extending a 
$[q,q-2]$ GRS code. Assuming Conjecture \ref{conj1} holds for $k=2$, i.e Conjecture ${\bf 1'}$  holds for  $m=q-2$, we deduce that $\mC^{\perp}$ is GRS. Thus $\mC$ is GRS.  \eep

Before, we state our next proposition justifying the remark above, we present another conjecture. This conjecture is clearly implied by the MDS conjecture. 
(The MDS conjecture states that the maximum length of a $k$-dimensional MDS code with $k<q$ is $q+1$ except when $q$ is even and $k=3,q-1$, in which cases it is $q+2$.) 
Since the MDS conjecture is widely believed, the same can be said about Conjecture \ref{rnc_mds_conj}. We will show that Conjecture \ref{conj1} implies Conjecture \ref{rnc_mds_conj}.

\begin{conjecture} [implied by the MDS conjecture]  \hfill \label{rnc_mds_conj} \\
There is no $[q+2,m]_q$ MDS code extending  a $[q+1,m]_q$ RS code, except when $q$ is even and $m=3, q-1$.\\
Equivalently,  the RNC in $PG(m-1,q)$ is a complete arc unless $q$ is even and $m=3, q-1$.
\end{conjecture}
 
\begin{proposition}    \label{conj12} Let $2 \leq m \leq q-2$. If Conjecture  ${\bf 1'}$ holds for $m$, then  Conjecture \ref{rnc_mds_conj} holds for $m$.
\end{proposition}
\bep  Let $\mC_1$ be a $[q+1,m]_q$ RS code generated by a matrix $G_m=  G_m(PG(1,q))$ as in  \eqref {eq:gen_GDRS}. Let $\mC$ be  $[q+2,m]_q$ MDS code generated by the matrix $[G_m \, |\, v]$ for some $v \in GF(q)^m$.  Let $\mC_2$ be the $[q+1,m]_q$ MDS code obtained by puncturing  $\mC$ on the $q+1$-th coordinate. We note that $\mC_2$ extends a $[q,m,\mD]_q$ RS code with $\mD = GF(q)$. Assuming Conjecture ${\bf 1'}$ holds for $m$, it follows that $v = a c_m(\infty)$ for some $a \in \GFm$, but then the last two colums of the matrix matrix $[G_m \, |\, v]$ are linearly dependent, contradicting the MDS property of $\mC$. Thus such a code  $[q+2,m]_q$ MDS code $\mC$ does not exist.
\eep
Proposition \ref{prop_rnc_mds} in the next section presents some of the known answers to Conjecture \ref{rnc_mds_conj}.

 Theorem \ref{main_thm} can be improved using results of Storme and Sz{\H{o}}nyi \cite{Storme_Szonyi}, \cite{Storme_Szonyi_1}. These results state that 
for $q$ odd sufficiently large and $4 \leq k \leq 0.09q +3.09$ any MDS extension of a $[n,k]$ GRS code with $n \geq (q+3)/2$ is GRS. Similarly 
for $q$ even sufficiently large, any MDS extension of a $[n,k]$ GRS code with $n \geq q/2+2$ if  $5 \leq k \leq 0.09q +3.59$  or $n \geq q/2+3$ if $k=4$,  
is GRS. Thus for such $n,k,q$ the generating functions of  deep holes of a  $[n,k]_q$ GRS code are as described in Theorem \ref{main_thm}.

\section{Reed Solomon codes of length $q+1$}  \label{gdrs}  
In this section $\mC$ is a $[q+1,k]_q$ RS code with evaluation set $PG(1,q)$. In terms of arcs, $\mC$ corresponds to a RNC in $PG(k-1,q)$. Let $G_k = G_k(PG(1,q))$ and $G_k^{\perp} = G_k^{\perp}(PG(1,q))$
denote the generator and parity check matrix of $\mC$ as given in  \eqref{eq:gen_GDRS}. As we showed  in Section \ref{mdsext}, the covering radius $\rho(\mC)$  satisfies:
\[  q-k \leq \rho(\mC)  \leq q+1-k.\]
It follows that $\rho(\mC)  =q+1-k \, \Leftrightarrow$  there exists a vector $u \in GF(q)^{q+1}$ at a distance of $q+1-k$ from $\mC \, \Leftrightarrow$  there exists a vector $v \in GF(q)^{q+1-k}$ such that the matrix $[G_k^{\perp} \,|\, v]$ generates a MDS code  $\, \Leftrightarrow$ the RNC in $PG(q-k,q)$ is an incomplete arc. This establishes the following theorem due to A.D{\"u}r.
\begin{theorem*} [D{\"u}r 1994 \cite{Dur1}] The covering radius of a $[q+1, k]_q$ RS code $\mC$ is $q-k$  if and only if  (any) RNC in $PG(q-k)$ is a complete arc. 
Equivalently there is no MDS extension of   $\mC^{\perp}$ by one digit.
\end{theorem*}
We can now restate Conjecture \ref{rnc_mds_conj} as follows: (where $k=q+1-m$)
\begin{conjecture*}  ${\bf 2'}$ 
The covering radius of a $[q+1,k]_q$ RS code is $q-k$ except when  $q$ is even and $k= 2, q-2$ in which cases it is $q+1-k$.
\end{conjecture*}
We present some of the known answers to Conjecture \ref{rnc_mds_conj}.
\begin{proposition}\label{prop_rnc_mds} Conjecture  \ref{rnc_mds_conj} is true for  \begin{enumerate}
\item (Roth and Seroussi \cite{Ro_Se})  $m=2$ and $3  \leq  m  \leq  \lfloor q/2 \rfloor  + 2$ except $m=3$ when $q$ is  even.
\item the exceptional cases $m=3, q-1$ with $q$ even.
\item (Segre \cite{Segre1}) $m=q-1$ with $q$ odd.
\item (Segre \cite{Segre1}) $m=q-2$ with $q$ odd.
\item (Segre \cite{Segre2}) $m=q-3$ with $q$ odd.
\item (Storme and Thas \cite{Storme_Thas_90}, Storme \cite{Storme_complete})  $\lfloor q/2 \rfloor  + 3  \leq  m  \leq q+3 - 6 \sqrt{q \ln q} $. 
\item (S. Ball \cite{Ball1}) any $m <q$ if $q$ is a prime.   
\item (S. Ball , \cite{Ball2})  $m  \leq 2 p  - 2$ where $q = p^h >p$ and $p$ is prime.
\item (Storme and Sz{\H{o}}nyi) \cite{Storme_Szonyi}) $4 \leq m \leq 0.09q+3.09$ with $q$ odd sufficiently large.
\item (Storme and Sz{\H{o}}nyi) \cite{Storme_Szonyi_1})  $q$ even sufficiently large, either $m=4$ or   $5 \leq m \leq 0.09q+3.59$.
\end{enumerate}
\end{proposition}
\bep \begin{enumerate}
\item  follows from the theorem of Roth and Seroussi above, together with the fact that for $m=2$ every MDS code is GRS.
\item  follows from the fact that for $q$ even, the matrices:
\beq  \label{eq:rsnucleus} H_3=[ c_3(x_1)\, |\,  \dots \,|\, c_3(x_q) \,|\, c_3(\infty) \,|\,  \mathcal N_3]\;\text{   and } \; H_{q-1}= [c_{q-1}(x_1) \, |\, \dots \, |\, c_{q-1}(x_q) \, |\,  \mathcal N_{q-1}  \, |\, c_{q-1} (\infty)] \eeq
 are parity check matrices of each other, and respectively generate a 
non-GRS $[q+2,3]$ MDS extension of a $[q+1,3]_q$ GRS code, and a non-GRS $[q+2,q-1]$ MDS  extension of a $[q+1,q-1]_q$ GRS code. 
\item Suppose there is a  $[q+2,q-1]$ MDS code.  Its dual $\mC$ is a $[q+2,3]$ MDS code. By the theorem of Segre \cite{Segre1}, it follows that the corresponding arc extends the RNC in $PG(2,q)$ contradicting the fact that the RNC in $PG(2,q)$ is a complete arc (Part 1).
\item  follows from Proposition \ref{conj12} and Proposition \ref{segre_prop}.
\item The result we need from \cite{Segre2} is that any $[q,3]$ MDS code is GRS for $q$ odd  (For a proof see \cite[Theorem 8.6.10]{Hirschfeld_book}). It follows that any $[q,q-3]$ MDS code is GRS for $q$ odd. Suppose the matrix  $[ G_{q-3}(PG(1,q)) \,| \, v]$ generates a $[q+2,q-3]$ MDS code. Puncturing on the first two coordinates gives  a $[q,q-3]$ MDS code,  which is GRS. The corresponding arc is thus contained in a RNC of  $PG(q-4,q)$. This arc  has $q-1$ points on the standard RNC. Since $q -1 = q-3+2$, once again appealing to the fact that 
a RNC in $PG(k-1,q)$ is uniquely determined by any $k+2$ points on it, it follows that $v$ also lies on the standard RNC. Thus  all $q+2$ columns of $[ G_{q-3}(PG(1,q)) \,| \, v]$ lie on the standard RNC in $PG(q-4,q)$ thus contradicting the fact that this matrix generates a MDS code.
\item, 9)-10) the RNC is complete in the range indicated as proved in the cited articles.
\item\!-8)  follow from the fact that the MDS conjecture holds in the parameter range indicated, as proved in the cited articles.

\end{enumerate} \eep

We now focus on the problem of classifying the deep holes of $\mC$ in  the exceptional cases of Conjecture ${\bf 2'}$.
For $q$ even, $k=2,q-2$, and $\mC$ a $[q+1,k]_q$ RS code,  the matrices $H_{q-1},H_3$ of \eqref{eq:rsnucleus}  generate MDS codes of length $q+2$ extending $\mC^{\perp}$.
Thus the theorem of D{\"u}r above implies that $\rho(\mC)$ is indeed $q+1-k$. 
We recall that the coset equivalence class of a deep hole $u$ is completely determined by its syndrome $S_{\mD}(u)$, and the equivalence class of $u$ completely determined by 
the projective syndrome $[S_{\mD}(u)]$.
\begin{theorem} \label{thm_gdrs_even} Let $\mC$ be a $[q+1,k]_q$ RS code, where $q$ is even and $k = 2, q-2$. Let $u$ be a deep hole of $\mC$.
\begin{enumerate}
\item If $k=q-2$:   the projective syndrome $[S_{\mD}(u)] = (0:1:0)$. Thus there is only one equivalence class of deep holes of $\mC$.
\item If $k=2$:  There is a bijective correspondence between the set of equivalence classes of deep holes of $\mC$ and the set of projective equivalence classes of 
 \emph{ordered hyperovals} of $PG(2,q)$.
  \end{enumerate}
\end{theorem}
\bep
Part 1): Here $k=q-2$.  Part 2) of the theorem of  Roth and Seroussi from the previous section states that  the code generated by $H_3$ is the only possible  MDS extension of a $\mC^{\perp}$.
Thus there is only one equivalence class of deep holes, represented by $[S_{\mD}(u)] = [0:1:0]$.\\ 

Part 2): Here $k=2$.  In this case the generator matrix is $G_2= \bbsm  1 & \hdots &   1&0  \\x_1  &  \hdots  & x_q  &1  \besm$, with $\{x_1, \dots, x_q\} = GF(q)$. Let $x_{q+1} = \infty$.
An \emph{ordered hyperoval} of $PG(2,q)$ is the ordered set of $q+2$ points of $PG(2,q)$ represented by a the columns of a generator matrix for a $[q+2,3]$ MDS code.
By a result of B. Segre (see \cite[Theorem 8.4.2]{Hirschfeld_book}), up to projective equivalence any such hyperoval is represented by a matrix
\[  G=\bbm  1 & \hdots &   1&0  &0\\x_1  &  \hdots  & x_q  & 1 & 0 \\ u_1 & \hdots &u_q & 0 &1 \bem \]
with the property that  $u_i=0$ if $x_i = 0, \infty$ and $u_i=1$ when $x_i=1$, and that $G$ generates a MDS code.
The condition that $G$ generates a MDS code can be stated as:  there are at most two zero entries of  $( bx_1+a -u_1, \dots, bx_q+a-u_q, b)$  for any $(a,b) \in GF(q)^2$. This is equivalent to $u$ being a deep hole of $\mC$. Since, the equivalence class of a received word $v \in GF(q)^{q+1}$ of $\mC$ (i.e. $\{ au +c : a \in \GFm, c \in \mC\}$) has a unique representative $u$ such that  $u_i=0$ if $x_i = 0, \infty$ and $u_i=1$ when $x_i=1$, it follows that equivalence classes of deep holes of $\mC$ are in bijective correspondence with projective equivalence classes of ordered hyperovals of $PG(2,q)$.  
\eep

The problem of classifying deep holes  of a $[q+1,2]$ RS code for $q$ even, is thus equivalent to the difficult problem of classifying hyperovals of $PG(2,q)$. (See Section 2 of \cite{Hirshfeld_Storme} for a survey of this problem).
The equivalence classes of deep holes $u$ are completely determined by their syndrome $[S_{\mD}(u)] \in PG(q-2,q)$. Thus, the hyperovals can be studied in terms of possible syndromes $S_{\mD}(u)$. This is done in the work of Storme and Thas  \cite{Thas_dual}. It is interesting to note that such a syndrome $[S_{\mD}(u)] = (a_0: \dots: a_{q-2})$ necessarily 
satisfies $a_0=a_2= \dots= a_{q-2}=0$.  (see Theorem 3.10 of  \cite{Thas_dual})\\

The problem of classifying deep holes of $\mC$ when $\rho(\mC) = q-k$ (for example Parts 1), 3)-5) of Proposition \ref{prop_rnc_mds}) is an open problem (since at least 1991,   see Remark 5 of \cite{Dur2}). By turning to the syndromes of the deep holes, and setting $m=q-k$,  this problem is equivalent to finding all points of $PG(m,q)$ which are not in the linear span of $m-1$ points of the standard RNC in  $PG(m,q)$.  We just consider the easiest case of this  problem.
\begin{theorem} \label{thm_gdrs_odd} For $k=q-2$ and $q$ odd,    $u=(u_1, \dots,u_{q+1})$  is a deep hole of $\mC$ if and only if its projective syndrome $[S_{\mD}(u)]$ does not lie on the standard RNC in $PG(2,q)$. Thus there are exactly $q^2$ equivalence classes of deep holes of $\mC$.
\end{theorem}
\bep Let $m = q-k = 2$. A point of $PG(2,q)$ which is not in the linear span of $m-1 = 1$ points of the standard RNC, is just a point which does not lie on the RNC.
\eep

\section{Classification of deep holes of RS codes of redundancy $3$} \label{red3}
In this section we will classify deep holes of $[n,k, \mD]_q$ RS codes $\mC$ of redundancy $n-k$ at most $3$. As remarked earlier, the cases $n-k$  being  $0$ and $1$ are uninteresting:
in the former case there are no deep holes, and in the latter case the deep holes are all  received words which are not codewords.
A generator and parity check matrix for $\mC$ is as given in  \eqref{eq:gen_GRS}, \eqref{eq:gen_GDRS}.
Since the projective syndrome $[S_{\mD}(u)]$ completely determines the equivalence class of a deep hole $u$, we will focus on determining the possible values  for $[S_{\mD}(u)]$. \\

First we consider redundancy $2$ case, i.e. $[k+2,k, \mD]_q$ RS code $\mC$ with $2 \leq k \leq q-1$. If $k=q-1$, then the length is $q+1$, and the theorem of D{\"u}r stated in Section \ref{gdrs}, together with the fact that the RNC in $PG(1,q)$ is complete (i.e. there are no $[q+2,2]$  MDS codes) implies that $\rho(\mC)=1$. Thus deep-holes of $\mC$ are those received words which are not codewords. For $k <q-1$, the length $k+2 \leq q$, and hence Proposition \ref{main_prop} implies that equivalence classes of deep holes of $\mC$ are in bijective correspondence with $[S_{\mD}(u)] \in PG(1,q)$ such that $[G_k^{\perp}(\mD) \,|\, S_{\mD}(u)]$ generates a $[k+3,2]$ MDS code. Since every $2$-dimensional MDS code is GRS, it follows that $[S_{\mD}(u)] \in PG(1,q) \setminus \mD$ \\

Now we  turn to RS codes of redundancy $3$.  Let $\mC$ be a $[k+3,k,\mD]$ RS code. Here $2 \leq k \leq q-2$. 
 We need some preliminary results and some notation. 
Let $\epsilon$ denote a fixed non-square element of $\GFm$ when $q$ is odd.
 The group $GL(2,q) = \{ \bbsm a & b\\ c & d \besm : ad - bc \neq 0\}$  acts on $GF(q)^2$ in the standard manner $v \mapsto gv$. This induces an action of the group $PGL(2,q) = GL(2,q) / \{ \pm \bbsm 1 & 0\\ 0 & 1 \besm \}$ on $PG(1,q)$ by $g \cdot x = (c+d x)/(a + bx)$. Here $x$ denotes $[c_2(x)]$.
Consider the action of $GL(2,q)$ on $GF(q)^3$   given by :
\beq g \cdot \xi =  \bbm a^2 & 2 a b & b^2\\  a c & a d + c b &   b d \\ c^2 & 2 c d & d^2 \bem \xi,   \quad  g=\bbsm a & b\\ c & d \besm, \; \xi \in GF(q)^3 \eeq
This induces an action of $PGL(2,q)$ on $PG(2,q)$ (see \cite[Proposition 2.5-2.6]{BGHK} for details). Under this action, it is easy to see that 
\[ g \cdot [c_3(x)] = [c_3( g \cdot x)]  \; \text{(which is $[c_3( \tfrac{c+d x}{a + bx})]$)}.\]
Since $PGL(2,q)$ acts transitively on $PG(1,q)$ it follows that $PGL(2,q)$ acts transitively on the standard RNC in $PG(2,q)$.
Thus the standard RNC  forms one orbit of the $PGL(2,q)$ action on $PG(2,q)$.
We also note that for $q$ even each element of $PGL(2,q)$ fixes the nucleus $(0:1:0)$, and thus this gives an orbit of size $1$.

\begin{lemma}  There are $3$ orbits for the action of $PGL(2,q)$ on $PG(2,q)$ given by: 
\begin{enumerate}
\item For $q$ even: i) the standard RNC, ii) the nucleus $(0:1:0)$, and iii) the orbit of $(1:0:1)$. \\The stabilizer of $(0:1:0)$ is $PGL(2,q)$, and the stabilizer 
of $(1:0:1)$ is
\[  G_{1} = \{ \bbsm 1+a & a \\ a & 1+a \besm : a \in GF(q)\} \; \text{isomorphic to the additive group of } \, GF(q)\]

We will denote the union of the two orbits i) and ii) by $\mO_1$. The orbit iii) will be denote $\mO_4$.
\item For $q$ odd: i) the standard RNC, ii) the orbit of $(0:1:0)$, and iii) the orbit of $(1:0:-\epsilon)$. \\The  stabilizer of $(0:1:0)$ is
\[G_0 = \{ x \mapsto a x^{\pm 1} : a \in \GFm\} \; \text{ isomorphic to the dihedral group of order $2(q-1)$} \]
The stabilizer of $(1:0:-\epsilon)$ is 
\[ G_{\epsilon}= \{ \bbsm a & b \\ c & d \besm : ||a+c \sqrt\epsilon||=1, (b,d)=\pm(\epsilon c, a)\},\]
isomorphic to the dihedral group of order $2(q+1)$. Here $|| \cdot|| : GF(q^2)^{\times} \simeq GF(q)[ \sqrt\epsilon]^{\times} \to \GFm$ is the norm.

We will denote  orbits i),ii) and iii) by $\mO_1, \mO_2$  and $\mO_3$ respectively.
\end{enumerate}
\end{lemma}
\bep We need to show that the orbits other than the RNC  for $q$ odd, and the RNC and its nucleus for $q$ even are as described.
Let $W$ denote the $3$-dimensional space of symmetric bilinear forms on  $GF(q)^2$. The group $GL(2,q)$ acts on $W$ by $(g \cdot B)(v,w) = B(g^{-1} v, g^{-1} w)$.
We consider a linear isomorphism:
\[ \Phi: GF(q)^3 \to W,  \;  \text{given by } \;  \Phi(M,N,P)(v,w) = v^T \bbsm P & -N \\ -N & M \besm w.\]   
The corresponding projective isomorphism will be also denoted $\Phi: PG(2,q) \to \bP W$. For later use, we record the formula
\beq \label{eq:Phi}  \Phi(M,N,P)((1,X)^T, (1,Y)^T) =  \frac{\text{det} \bbsm 1 & 1 & M \\X & Y & N\\ X^2 & Y^2 & P \besm }{Y-X} = MXY - N(X+Y) +P.\eeq
It is easy to check that $\Phi(g \cdot \xi) =    \text{det}(g)^2  \,g \cdot \Phi(\xi)$, and thus at the projective level $\Phi(g \cdot [\xi]) = g \cdot [\Phi(  \xi )]$.
The bilinear form $\Phi(M,N,P)$ is degenerate if and only if det$\bbsm P & -N \\ -N & M \besm = MP-N^2=0$. Thus  $\Phi$ carries  the orbit formed by the standard RNC   to the projective space of degenerate symmetric bilinear forms on $GF(q)^2$.  For the remaining orbits, it suffices to consider nondegenerate bilinear forms $B$. If $q$ is odd, it is well known that there exists $g \in GL(2,q)$ such that $(g \cdot B) (v,w)  =v^T \bbsm 0 &1 \\1 &0 \besm w$ or $v^T \bbsm 1 & 0 \\ 0 & -\epsilon \besm w$ depending on whether or not there is a nonzero vector $v$ with $B(v,v)=0$  (for example see \cite[Theorem 7.2.12]{Handbook}). Thus the orbits of $(0:1:0)$ and $(1:0: -\epsilon)$ are the other orbits. 
The stabilizer of $(0:1:0)$ and $(1:0:-\epsilon)$ are easy to compute and can also be found in  \cite[pp. 45-46]{Grove}.\\

For $q$ even, suppose $B(v,v)=0$ for all $v \in GF(q)^2$. In this case $B(v,w) = \alpha v^T \bbsm 0 &1 \\1 &0 \besm w$ for some $\alpha \in \GFm$ and thus $B = \Phi(0:1:0)$.
If there exists a vector $v$ such that  $B(v,v) \neq 0$, we may replace $v$  by $v/ \sqrt{B(v,v)}$ to achieve $B(v,v)=1$. The nondegeneracy implies there is a vector $w$ with $B(v,w)\neq 0$ and $B(w,w) \neq 0$. As above, we can assume $B(w,w)=1$. Thus $B = \Phi(1:0:1)$. The stabilizer of $(1:0:1)$ is clearly all matrices satisfying $g^Tg=\bbsm 1 &0\\0&1 \besm$, which is as described in the statement. The map  $\bbsm 1+a & a \\ a & 1+a \besm  \mapsto a$ is an isomorphism of $G_{1}$ with the additive group of $GF(q)$. \\
\eep

We define some  subsets of $PG(2,q)$ associated with an evaluation set $\mD \subset PG(1,q)$ of size $k+3$. 
\begin{itemize}
\item $\mathcal O_1(\mD)=\begin{cases} \{ [c_3(\delta)]: \delta \in PG(1,q) \setminus \mD\} &\text{ if $q$ is odd} \\
									\{ [c_3(\delta)]: \delta \in PG(1,q) \setminus \mD\} \cup (0:1:0) &\text{ if $q$ is even}. \end{cases}$

\item  $\mathcal O_2(\mD)= \{ \bar g \cdot (0:1:0) : \bar g \in PGL(2,q)/G_0,  x \neq y \in g^{-1} \cdot \mD \Rightarrow x \neq -y\}$  for $q$ odd.


\item  $\mathcal O_3(\mD)= \{ \bar g \cdot (1:0:-\epsilon) : \bar g \in PGL(2,q)/G_{\epsilon},  x \neq y \in g^{-1} \cdot \mD \Rightarrow x \neq \epsilon/y\}$  for $q$ odd.

\item  $\mathcal O_4(\mD)= \{ \bar g \cdot (1:0:1) : \bar g \in PGL(2,q)/G_{1},  x \neq y \in g^{-1} \cdot \mD \Rightarrow x \neq 1/y\}$  for $q$ even.
 \end{itemize}
We note that $\mO_i(\mD)$ is a subset of  $\mO_i$.  The notation of $\mO_i(\mD), i=2,3,4$ needs some explanantion.  The notation $PGL(2,q)/G_0$ stands for the left cosets of the stabilizer $G_0$ of $(0:1:0)$ in $PGL(2,q)$. Similarly for $G_{\epsilon}$ and $G_{1}$.
It is easy to show that if a subset $A$ of $PG(1,q)$ has the property that  for any pair of distinct elements $a, b \in A$,  i)  $a \neq -b$ ($q$ odd), or ii) $a \neq \epsilon/b$ ($q$ odd), or iii)
$a \neq 1/b$ ($q$ even), then 
for $g$ in i)  $G_{0}$, ii) $G_{\epsilon}$ , iii) $G_{1}$, the sets $gA$ also have the same property. Thus $\mO_i(\mD), i=2,3,4$ are well-defined.
 Moreover,  such a set $A$ has size at most i) $(q+3)/2$, ii) $(q+1)/2$, iii) $(q+2)/2$ respectively, by a simple application of pigeon hole principle.
 We recall the notation $\mD = \{x_1, \dots, x_{k+3}\}$. The size of these subsets can be expressed as:
\begin{IEEEeqnarray}{rCl}  \label{eq:size}
\nonumber  |\mO_1(\mD)| &=& q-k-2 \, \text{ if $q$ is odd, and } \, q-k-1 \, \text{ if $q$ is even}. \\ 
 |\mO_2(\mD)|  &= &| \{  \bar g \in G_0 \backslash PGL(2,q) : gx_i \neq - gx_j  \, \forall i \neq j\}|. \\
\nonumber |\mO_3(\mD)|  &= &| \{  \bar g \in G_{\epsilon} \backslash PGL(2,q) : gx_i \neq \epsilon/ gx_j  \, \forall i \neq j\}|. \\
\nonumber |\mO_4(\mD)|  &= &| \{  \bar g \in G_{1} \backslash PGL(2,q) : gx_i \neq 1/gx_j  \, \forall i \neq j\}|,
\end{IEEEeqnarray}
where $G_0 \backslash PGL(2,q)$  denotes  the set of right cosets of $G_0$ in $PGL(2,q)$. 
The exact values of the sizes of $\mO_i(\mD), i=2,3,4$ depends on the configuration of $\mD$ in $PG(1,q)$, and appears to be a hard problem.
\begin{theorem} \label{red3_thm} The the set of possible values of $[S_{\mD}(u)]$ is:
\begin{enumerate}
\item   $\begin{cases} (0:1:0)   &\text{ if $q$ is even,   $k=q-2$  }\\
\mO_2 \cup \mO_3  &\text{ if $q$ is odd,  $k=q-2$ } \end{cases}$
\item $\mO_1(\mD)$,  if $q-3 \geq k \geq \lfloor (q-1)/2 \rfloor$.
\item  $\mO_1(\mD) \cup \mO_2(\mD)$,  if $k=(q-3)/2$ with $q$ odd. 
\item $\mO_1(\mD) \cup \mO_2(\mD) \cup \mO_3(\mD)$,  if $2 \leq k \leq (q-5)/2$ with $q$ odd. 
\item  $\mO_1(\mD) \cup \mO_4(\mD)$,  if $2 \leq k \leq (q-4)/2$  with $q$ even. 
\end{enumerate}
\end{theorem}
\bep
 First we consider the length $q+1$ case (i.e. $k=q-2$). If $q$ is even, then the only possibility for $[S_{\mD}(u)]$ is the nucleus $(0:1:0)$ by Part 1) of Theorem \ref{thm_gdrs_even}. If $q$ is odd, then by Theorem  \ref{thm_gdrs_odd}, the possibilities for $[S_{\mD}(u)]$ is the complement of the standard RNC in 
$PG(2,q)$. This proves part 1).\\

Now we assume $k \leq q-3$. The length $k+3$ is then at most $q$ and  Proposition \ref{main_prop} implies that
$[ G_k^{\perp}(\mD) \, | \, S_{\mD}(u) ]$ generates a $[k+4,3]$ MDS code.  This always holds if $[S_{\mD}(u)] \in \mO_1(\mD)$. It remains to consider other possibilities for $[S_{\mD}(u)] $. For $q-3 \geq k  \geq  \lfloor (q-1)/2 \rfloor$, Theorem \ref{geom_thm} implies that 
there are no other possibilities. This proves part 2). \\

We now assume $2 \leq k \leq \lfloor (q-3)/2 \rfloor$.  Let $B(v,w)$ be the bilinear form on $GF(q)^2$ given by $\Phi(S_{\mD}(u))$. We are given that 
\beq \label{eq:B_cond} B((1,x_i),(1,x_j)) \neq 0,   \; \forall  x_i \neq x_j \in \mD. \eeq  
In case  $[S_{\mD}(u)]$ lies on the standard RNC or RNC $\cup$ its nucleus if $q$ is even, it follows that 
$[S_{\mD}(u)] \in \mO_1(\mD)$, which we have already considered.  Thus we assume  $[S_{\mD}(u)] \notin  \mO_1$.  If $q$ is even, that leaves us with  $[S_{\mD}(u)] \in \mO_4$. Writing  $[S_{\mD}(u)] = g^{-1} \cdot (1:0:1)$ for some  $g \in PGL(2,q)$, and let $g \cdot \mD = \{y_1, \dots, y_{k+3}\}$. It follows that:
\beq \label{eq:mo4}[ G_k^{\perp}(\mD) \, | \, S_{\mD}(u) ] = g^{-1} [ \mu_1 c_3(y_1) \,|\, \dots \,|\, \mu_{k+3} c_3(y_{k+3}) \,|\, (1,0,1)^T], \eeq
for some $\mu_1, \dots, \mu_{k+3} \in \GFm$. In this case the condition \eqref{eq:B_cond} is equivalent to  $y_i \neq 1/y_j$ for $i \neq j$. (It follows from \eqref{eq:Phi} that 
for $(M,N,P) = (1,0,1)$ the form $Mxy - N(x+y) +P = xy+1= xy-1$.) Hence  $[S_{\mD}(u)] \in \mO_4(\mD)$. This proves part 5).\\

Now we turn to the case $q$ odd, and $S_{\mD}(u) = (M,N,P)^T \notin \mO_1$. In case $(M,N,P) \in \mO_2$, let $(M,N,P)^T=g^{-1} \cdot (0,1,0)^T$ for some  $g \in PGL(2,q)$, and let $g \cdot \mD = \{y_1, \dots, y_{k+3}\}$. It follows that:
\beq \label{eq:mo2} [ G_k^{\perp}(\mD) \, | \, S_{\mD}(u) ] = g^{-1} [ \mu_1 c_3(y_1) \,|\, \dots \,|\, \mu_{k+3} c_3(y_{k+3}) \,|\, (0,1,0)^T], \eeq
for some $\mu_1, \dots, \mu_{k+3} \in \GFm$. In this case the condition \eqref{eq:B_cond} is equivalent to  $y_i \neq -y_j$ for $i \neq j$, because 
$Mxy - N(x+y) +P = -(x+y)$. Hence  $[S_{\mD}(u)] \in \mO_2(\mD)$. Similarly, if $(M,N,P) \in \mO_3$, we get $(M,N,P) \in \mO_3(\mD)$.
 As mentioned above the set $\mO_3(\mD)$ is  empty unless $k+3 \leq (q+1)/2$, thus for $k = (q-3)/2$, the possibility $S_{\mD}(u) \in \mO_3$ does not occur.
This proves parts 3)-4). 
\eep

We record the following theorem about canonical forms of non GRS $[n+1,3]$ MDS codes extending a GRS $[n,3]$ code. It will be useful to regard two codes $\mC, \mC'$ as diagonally equivalent if there is a diagonal Hamming isometry (a  diagonal matrix in $GL(n,q)$) which carries $\mC$ to $\mC'$. Note that diagonally equivalent codes are monomially equivalent but the converse is not true in general.  At the level of arcs, diagonal equivalence  yields the notion of ordered arcs, where as monomial equivalence yields the 
the notion of  (unordered) arcs.
\begin{theorem} \label{thm_canon} Let $\mC$ be a non GRS $[n+1,3]$ MDS code extending a $[n,3]$ GRS code $\mC_1$ where $n \geq 5$. Up to diagonal equivalence, $\mC$ is the code generated by one of the families of matrices  $M_1, M_2, M_3$ below. 

Equivalently let $\mathcal A$ be an ordered $n+1$-arc in $PG(2,q)$ with the first $n$ points (but not the last) on a RNC (where $n \geq 5$), then $\mathcal A$ is projectively equivalent to the ordered arc defined by the columns of one of the families of matrices $M_1, M_2, M_3$ below. 

 In the following, $\mD =  \{x_1, \dots, x_n\} \subset PG(1,q)$ denotes  a subset of $n \geq 5$ distinct points satisfying certain conditions.
\begin{enumerate}
\item   $\mD$ satisfies $ x_i \neq - x_j$   if  $i \neq j$. In this case $n \leq (q+3)/2$ if $q$ is odd and $n \leq q+1$ if $q$ is even.
\beq M_1 = \bbm 1 & \ldots& 1 & 0\\x_1 & \ldots &x_n & 1 \\ x_1^2 & \ldots & x_n^2 & 0 \bem \eeq
\item $q$ is odd, $n \leq (q+1)/2$, and  $\mD$ satisfies $ x_i \neq \epsilon/ x_j$   if  $i \neq j$.
\beq M_2 = \bbm 1 & \ldots& 1 & 1\\x_1 & \ldots &x_n & 0 \\ x_1^2 & \ldots & x_n^2 & -\epsilon \bem \eeq
\item $q$ is even, $n \leq (q+2)/2$, and  $\mD$ satisfies $ x_i \neq 1/ x_j$   if  $i \neq j$.
\beq M_3 = \bbm 1 & \ldots& 1 & 1\\x_1 & \ldots &x_n & 0 \\ x_1^2 & \ldots & x_n^2 & 1 \bem \eeq
\end{enumerate}
\end{theorem}
\bep From the fact that a RNC in $PG(2,q)$ is uniquely determined by any $5$ points on it, it follows that the matrices $M_i$ above do not generate a GRS code for $n \geq 5$ (the corresponding arcs do not lie on a RNC).  To prove that the code $\mC$ in question is diagonally equivalent to the code generated by one of the matrices of the type $M_i$,
let $C_1$ be diagonally equivalent to the code generated by a matrix $G = [c_3(t_1) \,|\, \dots \,|\, c_3(t_n)]$. Thus there is a vector $v \in GF(q)^3$ such that 
$[G \,|\, v]$ generates the non-GRS code $\mC$. The analysis of such matrices $[G \,|\, v]$ was carried out in the proof of Theorem \ref{red3_thm} (see \eqref{eq:mo4}, \eqref{eq:mo2}). It was shown that there are matrices $P \in GL(3,q)$ and a diagonal matrix $Q \in GL(n+1,q)$ such that $P   [G \,|\, v] Q$ is of the type $M_1, M_2$ or $M_3$.
In other words $\mC$ is diagonally equivalent to the code generated by one of the types of matrices $M_i$.
\eep
We note that two distinct matrices of the type, say $M_2$ may represent the same MDS extension $\mC$ of $\mC_1$. In order to count the diagonal equivalence classes of codes
$(\mC_1, \mC)$ where $\mC$ is a  $[n+1,3]_q$  MDS and non GRS code extending a $[n,3]_q$ RS code  $\mC_1$, we have to factor out the left action of $G_0, G_{\epsilon}, G_1$ on generator matrices of the type $M_1, M_2, M_3$.  It is convenient to use the language of arcs. We will now count the number of 
projective equivalence classes of  ordered arcs $(\mathcal A_1, \mathcal A)$ where $\mathcal A$ is an ordered $n+1$-arc not contained in  a RNC, but its first $n$ points form the arc $\mathcal A_1$ which is contained in a RNC.  Let $\mathcal M_i$ be the set of ordered arcs (without using projective equivalence) arising from matrices of the type $M_i$. Let $\mG_i \subset PGL(2,q)$ be the stabilizer of the point represented by the last column. It is easy to see that $\mG_i$ acts freely (i.e. without fixed point) on $\mathcal M_i$. This is because the  only element of $PGL(2,q)$ which fixes $3$ points is the identity transformation.   The quotient $\mG_i \backslash \mathcal M_i$ gives the projective equivalence classes of ordered arc pairs $(\mathcal A_1, \mathcal A)$ that we are trying to count and which are of type $M_i$.
It is straightforward to count the relevant quantities:
$|\mathcal M_1|=(q+1)! /  (q+1-n)! $ if $q$ is even, and  
\[ |\mathcal M_1| = \frac{\tfrac{q-1}{2} ! \, 2^n}{(\tfrac{q-1}{2}-n)!} + \frac{\tfrac{q-1}{2} ! \, 2^n\, n}{(\tfrac{q+1}{2}-n)!}+
\frac{\tfrac{q-1}{2} ! \, 2^{n-2}\, n(n-1)}{(\tfrac{q+3}{2}-n)!} \quad \text{ if $q$ is odd}.\]
Here we use the convention $(-m)!=\infty$ for natural numbers $m$. We illustrate the method we use to obtain $|\mathcal M_1|$ for $q$ odd. The other cases are similar.
We may write $PG(1,q)$ as the disjoint union of $(q+3)/2$ sets of the form $ \{\infty\}, \{0\}, \{\pm \alpha_1\}, \dots , \{\pm \alpha_{(q-1)/2}\}$. We note that $\mathcal M_1$ consists of 
$n$-tuples $(z_1, \dots,z_n)$ such that we pick at most one element from each of  the $(q+3)/2$ sets above.
By similar methods, we obtain  
\begin{IEEEeqnarray*}{rCl}
 |\mathcal M_2| &=& (\tfrac{q+1}{2} ! \, 2^n )/ (\tfrac{q+1}{2}-n)! \\
 |\mathcal M_3| &=& (\tfrac{q}{2} ! \, 2^n)/ (\tfrac{q}{2}-n)!+ (\tfrac{q}{2} ! \, 2^{n-1} \,n)/(\tfrac{q+2}{2}-n)!
\end{IEEEeqnarray*}
The groups $\mG_i$ have been computed previously: $\mG_1$ is $PGL(2,q)$ if $q$ is even and isomorphic to a dihedral group of order $2(q-1)$ for odd $q$.
The group $\mG_2$ isomorphic to a dihedral group of order $2(q+1)$, and the 
The group $\mG_3$ isomorphic to the additive group $(GF(q),+)$.
Thus we obtain that the number of ordered arc pairs $(\mathcal A_1, \mathcal A)$  of the type $M_i$ equals:
\begin{enumerate}
\item $(q-2)! / (q+1-n)!\quad$  if $i=1$ and  $q$ is even. Here $ n\leq q+1$.\\
\item $ \tfrac{q-3}{2} !  \, 2^{n-4}\, \left[ (q+1)(q+3-2n)  + n(n-1)\right]/  (\tfrac{q+3-2n}{2} \,!)\quad$ if $i=1$ and  $q$ is odd. Here $n \leq (q+3)/2$\\
\item $\tfrac{q-1}{2} ! \, 2^{n-2} /  (\tfrac{q+1-2n}{2} \, !)\quad$ if $i=2$. Here $n \leq (q+1)/2$.\\
\item $\tfrac{q-2}{2} ! \, 2^{n-2} \, (q+2-n) / (\tfrac{q+2-2n}{2} !) \quad$ if $i=3$. Here $n \leq (q+2)/2$.
\end{enumerate}

\section{Conclusion}
We solve the problem of classifying deep holes of $[n,k]_q$ RS codes for $k \geq (q-1)/2$ for non prime $q$, which was posed as an open problem in the concluding remarks of \cite{ZCL}. The problem for $k < (q-1)/2$ is open. We solve the problem for $n=k+3$ and all $k$. We also solve the problem for $k=2, n=q$  with $q$  odd, by reducing it to Segre's `oval equals conic' theorem.  For $k=2, n=q+1$ with $q$ even, we show that the problem is equivalent to the difficult problem of classifying hyperovals in projective planes. 
Finally, we obtain canonical forms for $[n+1,3]_q$ MDS but non-GRS codes extending a $[n,3]_q$ GRS code.
\bibliographystyle{IEEEtran}
\bibliographystyle{plain}
\bibliography{refs}
\nocite{}
\end{document}